\documentclass[conference]{IEEEtran}
\IEEEoverridecommandlockouts
\usepackage{amsmath,amssymb,amsfonts}
\usepackage{subfigure}
\usepackage{slashbox}
\usepackage{array}
\usepackage{pifont}
\usepackage{multicol}
\usepackage{textcomp}
\usepackage[utf8]{inputenc}
\usepackage{textcomp}
\usepackage[utf8]{inputenc}
\usepackage{graphicx}
\usepackage{hyperref}
\usepackage{mathtools} 
\usepackage{extarrows}
\usepackage{amsmath}
\usepackage{amsfonts}
\usepackage{marvosym}
\usepackage{amssymb}
\usepackage[pscoord]{eso-pic}
\def\BibTeX{{\rm B\kern-.05em{\sc i\kern-.025em b}\kern-.08em
    T\kern-.1667em\lower.7ex\hbox{E}\kern-.125emX}}
    
\usepackage{fancyhdr}

\AddToShipoutPictureFG*{\AtPageUpperLeft{\put(250,-18){\makebox[\paperwidth][l]{\textbf{\begin{large} \textit{Journal of Metaverse} \end{large} }}}}}
\AddToShipoutPictureFG*{\AtPageUpperLeft{\put(275,-30){\makebox[\paperwidth][l]{\textbf{Review Article}}}}}

\AddToShipoutPictureFG*{\AtPageUpperLeft{\put(20,-40){\makebox[\paperwidth][l]{\textbf{Received:} 2022-02-11 $\vert$ \textbf{Reviewing: }2022-02-14 \& 2022-03-31 $\vert$ \textbf{Accepted:} 2022-03-31 $\vert$ \textbf{Online:} 2022-03-31 $\vert$ \textbf{Issue Date:} 2022-06-30}}}}
\AddToShipoutPictureFG*{\AtPageUpperLeft{\put(210,-50){\makebox[\paperwidth][l]{\textbf{Year:} 2022, \textbf{Volume:} 2, \textbf{Issue:} 1, \textbf{Pages:} 8-15 }}}}

\begin{document}

\pagestyle{fancy}
\fancyhf{}
\fancyhead[C]{\textbf{\begin{large} \textit{Journal of Metaverse} \end{large}}}
\makeatletter
\renewcommand{\headrulewidth}{0pt}

\title{Applying Digital Twins in Metaverse: User Interface, Security and Privacy Challenges}

\author{\IEEEauthorblockN{1\textsuperscript{st} Saeed Banaeian Far$^*$}
\IEEEauthorblockA{\textit{Department of Electrical Engineering,}\\ \textit{Yadegar -e- Imam Khomeini (rah),  Shahr-e-rey Branch,} \\
\textit{Islamic Azad University,  Tehran, Iran}\\
saeed.banaeian.far@gmail.com\\ 0000-0003-2520-8492}
 \and 
\IEEEauthorblockN{2\textsuperscript{nd} Azadeh Imani Rad}
\IEEEauthorblockA{\textit{Department of Electrical Engineering,}\\ \textit{Yadegar -e- Imam Khomeini (rah),  Shahr-e-rey Branch,} \\
\textit{Islamic Azad University,  Tehran, Iran}\\
azadeh\_{imany}@yahoo.com\\ 0000-0002-8532-9379}

\thanks{$^*$Corresponding author: Saeed Banaeian Far - \textit{This article has been accepted in} "Journal of Metaverse". \textbf{You can cite as (APA):} Banaeian Far, S. \& Imani Rad, A. (2022). Applying Digital Twins in Metaverse: User Interface, Security and Privacy Challenges. Journal of Metaverse, 2 (1), 8-16. Retrieved from \url{https://dergipark.org.tr/en/pub/jmv/issue/67967/1072189}  }
}

\maketitle

\begin{abstract}
Digital Twins (DTs) are a conventional and well-known concept, proposed in $70$s, that are popular in a broad spectrum of sciences, industry innovations, and consortium alliances. However, in the last few years, the growth of digital assets and online communications has attracted attention to DTs as highly accurate twins of physical objects. Metaverse, as a digital world,  is a concept proposed in $1992$ and has also become a popular paradigm and hot topic in public where DTs can play critical roles. This study first presents definitions, applications, and general challenges of DT and Metaverse. It then offers a three-layer architecture linking the physical world to the Metaverse through a user interface. Further, it investigates the security and privacy challenges of using DTs in Metaverse. Finally, a conclusion, including possible solutions for mentioned challenges and future works, will be provided. 
\end{abstract}


\section{Introduction}
\label{intro}
Recently, the rapid growth of the internet and digital communications has increased the popularity of digital home-based and remote jobs. Consequently, and also as a result of increased internet-based communications, hackers and malicious users are motivated to commit fraud against regular users \cite{intsec}. Therefore, the demand for security and privacy on the internet has increased in the last two decades. 

Visual effects in internet-based communication have made them more attractive and efficient. Online shops and virtual meetings allow people to do their outdoor businesses in their homes efficiently, fast, and with lower expenses \cite{visual}. Designing user-friendly and easy-to-use services attracts customers. Therefore, service providers try to improve their services to provide better visual effects and user-friendly designs. 

High-level and accurate simulation create natural feelings for users and greatly assists technology \cite{design}, and enables designers to predict future effects of products and prevent possible risks. Digital twins (DTs) provide the most realistic simulation of physical objects in a way that they can accurately indicate and predict all the physical output of the computer \cite{dtdef}. Highly accurate DTs greatly help the industry and protect physical objects (for more details, refer to Section \ref{dtsection}). Creating DTs has been a state-of-the-art field of science and technology for many years (and will be in the future).  

Metaverse, as a concept proposed in $1992$, has become a popular blockchain-based concept/technology in public after renaming \textit{Facebook} to \textit{Meta} (for more detail about blockchain and Metaverse, refer to Sections \ref{bcsection} and \ref{mvsection} and \cite{bitcoin,ijns,mvmv1,mvmv2}). However, it has many regulatory, security, and privacy gaps that should be solved. It is believed that applying DT designing ideas in Metaverse can create natural/actual fillings in Metaverse-based digital things for users and make Metaverse more attractive and user-friendly. Virtual Reality (VR) headsets and Augmented Reality (AR) are two well-known instruments and technologies linking users to the digital world \cite{arvr}. In addition to VR and AR, Artificial Intelligence (AI) and Machine Learning (ML) are two fields of science and technology that greatly develop Metaverse and virtual spaces \cite{aiml}. 

\subsection{Contribution}
This paper consists of some parts as below\footnote{This paper can be assumed as a review paper consisting of a contribution and discussion.}:
\begin{itemize}
\item \textit{Review:} This study presents a review of the Metaverse and related concepts. 
\item \textit{Contribution:} This study presents a three-layer architecture containing a user interface layer for linking the physical world to the digital one (Metaverse).
\item \textit{Discussion:} As the digital world can be approximately equivalent to the physical world, it is believed that DTs can play critical roles in Metaverse. Therefore, this study discusses the security and privacy challenges of applying DTs in the Metaverse. 
\item \textit{Possible solutions and future works:} This study presents possible solutions for some of the discussed challenges and suggests future works in this field. 
\end{itemize}

\subsection{Outline}
The rest of the paper is organized as follows: Section \ref{defapp} presents the definitions, applications, and general challenges of preliminaries concepts. Section \ref{combdtmv} offers the architecture of combining DTs with Metaverse. Section \ref{secchall}, as the paper's most important section, discusses the security and privacy aspects of applying DTs in Metaverse. Finally, Section \ref{conc} concludes this study and presents some possible solutions and future works.

\section{Definitions, Applications, and General Challenges of Preliminaries} \label{defapp}
In this section, the paper's preliminaries general definitions, applications, and challenges are presented.

\subsection{Blockchain} \label{bcsection}
\begin{figure}[h]
\centering
\includegraphics[scale=0.42]{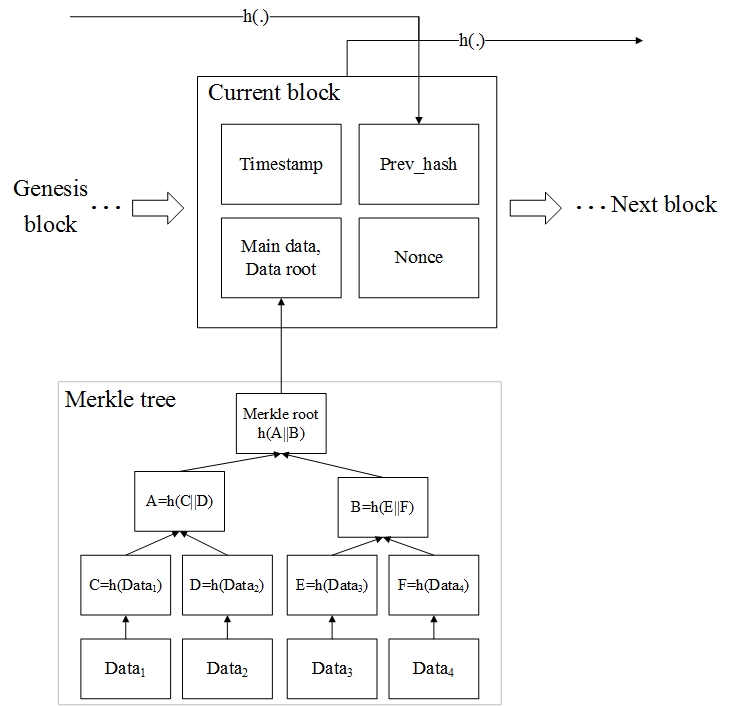}
\caption{The Blockchain Structure \cite{bitcoin,ijns}}
\label{bcstr}
\end{figure}

Blockchain was practically proposed by an anonymous author named S. Nakamoto as Bitcoin infrastructure \cite{bitcoin}. Blockchain technology is a distributed computer-based ledger that provides immutability, transparency, and autonomy \cite{ijns}. The blockchain structure is illustrated in Fig. \ref{bcstr}.

As shown in  Fig. \ref{bcstr}, blockchain's blocks consist of four main blocks of \textit{Timestamp} (current block's generating time), \textit{Prev\_{hash}} (previous block's hash), \textit{Nonce}, \textit{Main data} as the recorded in the current block, and \textit{Data root} as the calculated Merkle tree root \cite{mt} related to the main data.  

\subsubsection{Application}
In the first few years, blockchain technology was used as infrastructure for cryptocurrencies (especially Bitcoin). More recently, it has been applied in financial transactions infrastructure and as a tool in money-laundry by criminals (the idea of private cryptocurrencies was not meant to assist criminals. Unfortunately, however, private cryptocurrencies are a popular tool among them). However, blockchain is a well-known technology with numerous applications in industry, science, and state-of-the-art research \cite{bcapp} and has also attracted governments. 

Blockchain applications are not limited to financial transactions, and blockchain technology can be applied as infrastructure for $i)$ storage, $ii)$ users are not trusted, $iii)$ transparency, $iv)$ immutability, $v)$ peer-to-peer connection, and $vi)$ accessibility. Fields that require these properties include digital healthcare records, insurance, smart grids, internet of things (IoT) and industrial IoT, reporting, rewarding, payment, and reputation systems.

\subsubsection{General Challenges}
Although blockchain technology is well-known and popular, it comes with significant challenges, including scalability, security, energy and cost, latency and complexity, and regulation and government. Some of these challenges have been solved, and others have various solutions. However, no specific universal scheme exists for all. 

\subsection{Non-fungible Token}
\begin{figure}[h]
\centering
\includegraphics[scale=0.72]{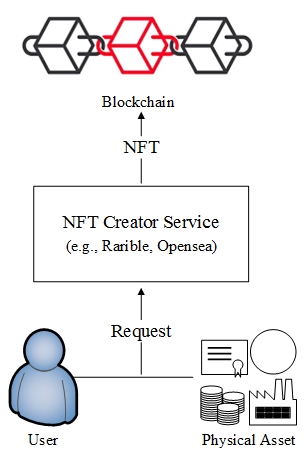}
\caption{Non-fungible Token Diagram}
\label{nftfig}
\end{figure}

The concept of the non-fungible token (NFT) first arose from the Ethereum token standard in $2017$ \cite{nft1}. It was proposed to distinguish between submitted tokens by distinguishable signatures. The NFTs are transacted as valuable digital assets on public blockchains (e.g., Ethereum blockchain). The uniqueness of NFTs enables them to link themselves to particular identities or digital assets. Additionally, this feature was considered for decentralized applications (DApp). The process of generating NFT is concisely illustrated in Fig. \ref{nftfig}.

Using NFTs as distributed ownership documents is the best choice for digital things' proof of ownership (e.g., digital vehicles, lands, markets, movies, houses, etc) in the Metaverse. 

\subsubsection{Application}
The first application of NFTs was linking them to artworks, and music albums \cite{nft2}. Recently, NFTs have attracted the attention of investors in digital fields so that they are investing massive values in this field. As a new and hot topic in the investigation, NFTs are used to transfer digital lands in Metaverse (Metaverse will be described in Section \ref{mvsection}), where Decentraland and Sandbox projects are the most famous.

\subsubsection{General Challenges}
As digital assets, NFTs have various challenges \cite{nft3}. Although uncertainty in prices is the most significant, other challenges exist, including proof of uniqueness, buyer and seller security, regulatory issues, cyber-attacks, evaluation, and money-laundering.

\subsection{Metaverse} \label{mvsection}
Assume a computer-based or virtual environment where one can find all physical things, services, friends and family, buildings, world map, and the Universe there.  Metaverse, defined in $1992$, consists of the two phrases of \textit{Meta} and \textit{Universe}, which provides a 3D virtual world that tries to present an approximately equal simulation of the physical world \cite{mv1}. Several newly-established companies (e.g., Decentraland, Sandbox, Upland, etc) and many famous active companies in information technology (Facebook or Meta, Microsoft, Google, Samsung, etc) focus on Metaverse, and they try to release their Metaverses as new services.  

In $2021$, Duan \textit{et al.} proposed a general three-layer architecture for Metaverse \cite{mv2}. In this proposed general model of Metaverse, the Interaction layer links the Ecosystem and Infrastructure. Based on the proposed architecture by Duan, the mentioned seven layers of Metaverse are summarized in the three phases below:
\begin{enumerate}
\item \textit{Infrastructure:} Fundamental and physical requirements, including blockchain, network, and computational powers, are established in this layer.  
\item \textit{Interaction:} This layer connects the Ecosystem and Infrastructure layers, and the contents of Metaverse are created in this layer. 
\item \textit{Ecosystem:} It is the parallel digital world or Metaverse. This layer involves user-generated content, economics, and AI. 
\end{enumerate}

\begin{figure}[h]
\centering
\includegraphics[scale=0.47]{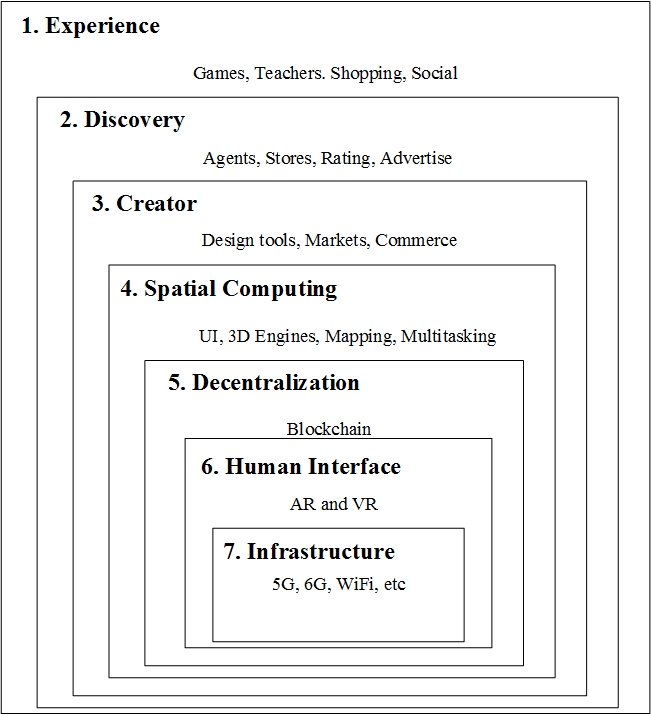}
\caption{Metaverse Architecture and Main Layers \cite{mv1,mv3}}
\label{metlayfig}
\end{figure}

As shown in Fig. \ref{metlayfig}, Metaverse architecture is defined in seven main layers \cite{mv1,mv3} (these seven layers could be assumed as the equal architecture, with more details, of the three above layers). These layers are described in the following.

\begin{enumerate}
\item \textit{Experience:} This is the closest layer to users in the physical world, where it could be assumed equal to the application layer in network architectures.   
\item \textit{Discovery:} This layer is driven by creators and service providers for motivating and informing users/communities. This layer consists of the required information, including related content, live streams, advertising emails and messages, and notifications, which are broadcast/informed by the creators' marketing departments.  
\item \textit{Creator:} Creators, who power the previous layer, are present in this layer. They design, create and develop their applications for end-users.  
\item \textit{Spatial Computing:} This layer supports a hybrid form of computation that reduces the boundaries between the physical and the digital worlds. This layer can be assumed the backbone of the creator layer, consisting of 3D engines (for showing geometry and animation), mapping and Interpreting, spatial mapping, integration of data from sensors, and user interfaces.  
\item \textit{Decentralization:} Distributed computing is an the essential primary in Metaverse, which provides a flexible ecosystem for developers and reliability for users. Blockchain technology plays a critical role in this layer as the essential component that supports decentralized infrastructure and is responsible for queries.  
\item \textit{Human Interface:} Physical-to-digital and digital-to-physical translators are present in this layer to make sense of the digital world and create a natural feel for users based on the digital world. In addition to AR and VR, smart glasses, 3D printers/scanners, biosensors, and perhaps even customer neural could be physical-to-digital and digital-to-physical translators.  
\item \textit{Infrastructure:} This layer, named the Internet layer, allows users and their devices to connect to the digital world. Even though 6G further improves speeds, 4G, 5G, and WiFi are famous examples of this layer, and Web 3.0 is the best choice for Metaverse.  
\end{enumerate}

\subsubsection{Application}
Based on the definition of Metaverse its applications are easy to guess. All daily needs are digitally supported since it is the DT of the same physical world. Additionally, users can also have real senses if they have VR and AR instruments \cite{mv3}. Therefore, users can resolve their daily needs on Metaverse.

High-level Metaverse applications include  military applications (on access to Tactical Augmented Reality (TAR)), real estate applications (on access to VR),  manufacturing applications, education applications (on access to VR headset), travel, shopping, virtual meetings, and conferences.

Practical examples of Metaverse applications include film producers showing their film trailers in Metaverse, fashion show companies providing their showplaces in Metaverse, online markets selling their products online supporting real-senses, holdings setting commercial meetings, and game producers presenting their creations on Metaverse. 

\subsubsection{General Challenges}
As with traditional social networks, Metaverse has several challenges \cite{mv3}. However, decentralized architecture, where no authority nor regulations prevail, comes with more challenges (more descriptions for the following challenges are not provided, and challenges are only reviewed in this section since they are approximately similar to those mentioned in Section \ref{secchall}, and you can find them there with detailed descriptions). 

Examples of Metaverse challenges are issues with reputations and identities, data security, money-laundering, currency (cryptocurrency) security and payments, regulation, judgment, legality, ownership proof (e.g., of data, NFT, DT, etc), global time, misbehaving detection, and usefulness for criminals.

\subsection{Digital Twin} \label{dtsection}
\begin{figure}[h]
\centering
\includegraphics[scale=0.38]{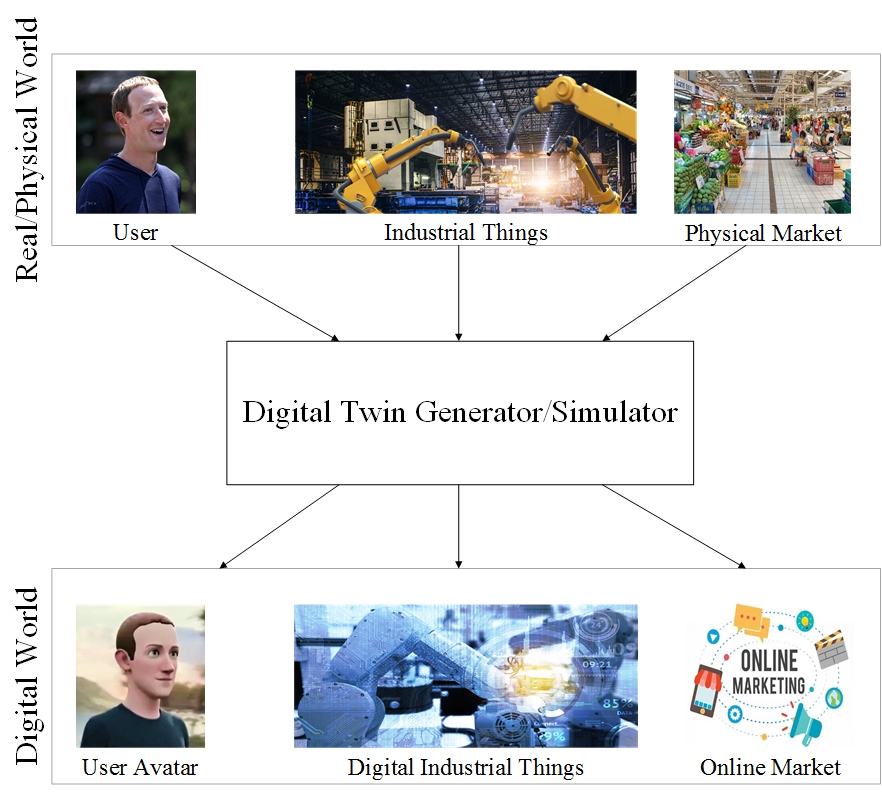}
\caption{The Connection Between Physical and Digital Objects}
\label{dtfig}
\end{figure}

DT is a virtual model of a process, product, or service proposed by NASA in $1970$. Based on the input data, DTs provide process prediction and risk prevention in the physical world \cite{dtdef}. These two main achievements enable managers to have well-organized plans for maintaining their products from possible risks and present better information about them to customers. Therefore, managers and customers have accurate details of the products to get the highest efficiency and adoption. As shown in Figs. \ref{dtfig} and \ref{rel}, DTs are generated/simulated by computers, 3D scanners, and developers based on the original physical objects. 

As an example and for more clarification, for generating the digital twins of the real world, six steps, including $i)$ 2D paper (map), $ii)$ 2D Digital map, $iii)$ 3D Digital map, $iv)$ High definition (HD) live map, $v)$ Indoor city, and $vi)$ Digital twin of the real world, should be passed.

\subsubsection{Application}
Developers and researchers try to create more accurate DT as they have numerous applications in the industry, science, and academic research \cite{dtapp}. DTs enable operators to predict the future condition of instruments and prevent possible risks. Additionally, DTs show the effect of real things (products) and simulate their behavior in different environments so that the company owners are at no risk (e.g., physical and financial). DTs empower smart vehicles, electronic healthcare systems, IoT, IIoT, and Industry $4.0$ to improve their output and efficiency (it should be noted that DTs' applications are not limited to the mentioned items). 

In practice, DT  applications in science and technology reduce production time and help in preparing the final products, retail market modeling for getting more customers, climate prediction, and many fields in business.

\subsubsection{General Challenges}
As DT challenges \cite{dtdef,dtchall1,dtchall2} are a critical part of this paper, they will now be itemized in the following:
\begin{itemize}
\item Data analytic challenges within the field of machine and deep learning 
\item IoT and IIoT challenges
\item Interoperability (composability, scalability, heterogeneity)
\item Security (integrity, confidentiality, availability)
\item Dependability (reliability, maintainability, availability, safety)
\item Sustainability (adaptability, resilience, reconfigurability, efficiency)
\item Reliability (robustness, predictability, maintainability)
\item Predictability (accuracy, compositionality)
\item signal processing related challenges 
\item Latency in real-time communication  
\item Large computations, data volume, data generation rate,  variety of data, veracity of data, and fast archival retrieval
\item Data management 
\item Ethical, legal, and societal issues
\item Blockchain adoption
\end{itemize}

\section{The Combination of Digital Twins and Metaverse} \label{combdtmv}
\begin{figure*}[h]
\centering
\includegraphics[scale=0.34]{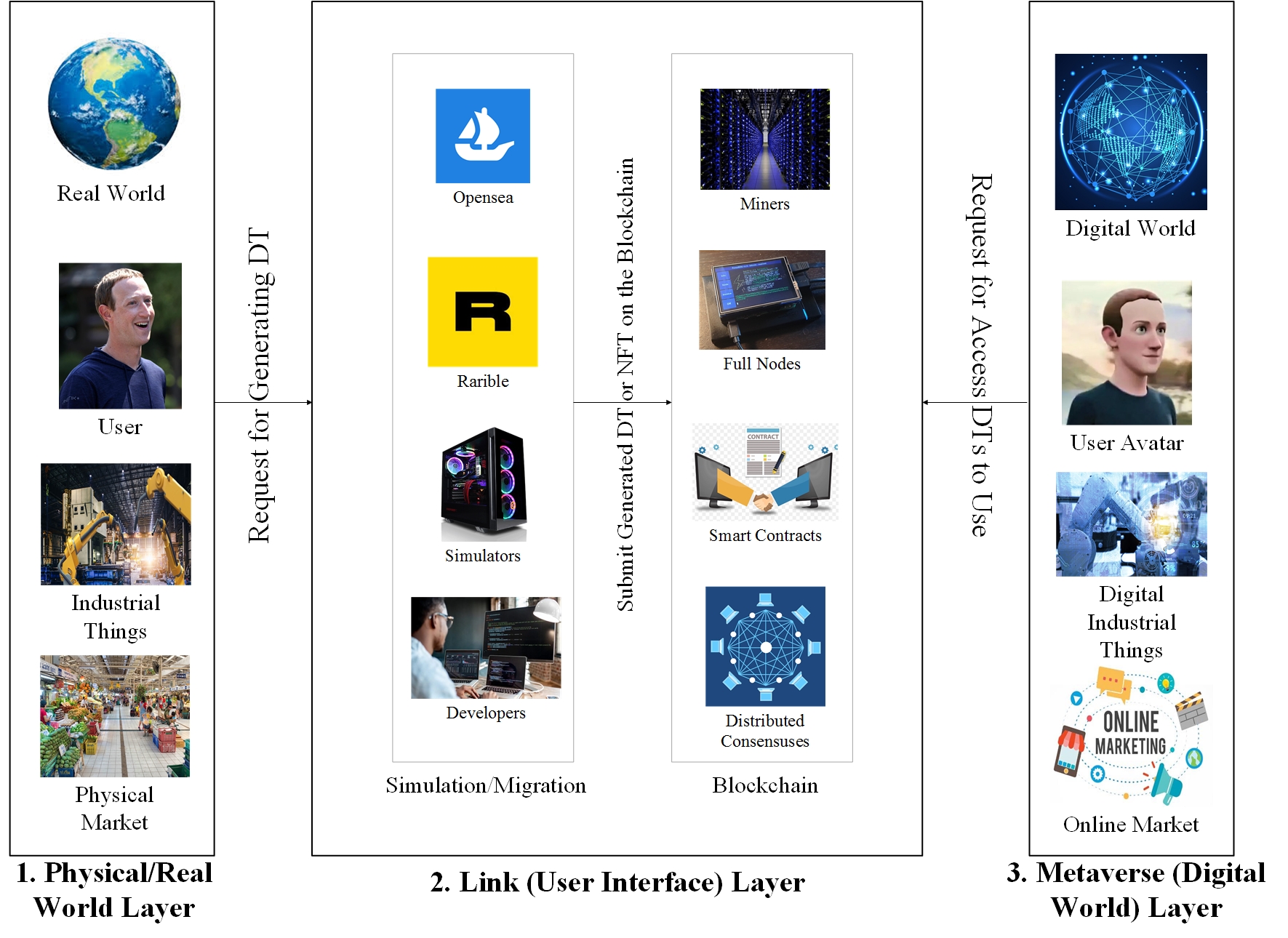}
\caption{The Three-layer Architecture of Connection between Physical World and Metaverse with the Approach of Applying DTs in Metaverse}
\label{rel}
\end{figure*}

This section suggests a three-layer architecture indicating the relationship between DTs and Metaverse, empowered by blockchain technology supporting a Metaverse interface. Assumed blockchain-based NFT is equal to DT has various benefits \cite{dtben} (Fig. \ref{rel} clarifies this relation). The benefits of using DTs in Metaverse are listed in the following.

\begin{itemize}
\item \textit{Immutability and transparency for DTs transactions:} Based on the above assumption, immutability and transparency are provided in the DTs transactions, including buying, selling, or ownership transfer. It, therefore, can be said that they are protected against cyber-frauds. 
\item \textit{Automation:} Blockchain supports autonomy so that no authority or privileged insider can interfere in DTs results. As a result, Metaverse-based DTs results are reliable. 
\item \textit{DTs identity and legitimacy:} Based on decentralized management in Metaverse, all allowed identities, especially of DT, are legal since they all are accepted under a consensus protocol. 
\item \textit{Security and reliability:} Blockchain solves some security challenges (not all). Therefore, Metaverse-based DTs are more secure and reliable than their centralized counterparts. 
\item \textit{High accuracy tracking for DTs globally:} Blockchain properties, including linking blocks, transparency, and immutability, provide global traceability for DTs and the related correspondences.  
\item \textit{Safeguarding product lifecycle:} As with the previous item, each Metaverse-based DT and its linked real-world product lifecycle are easily controllable. 
\item \textit{Peer-to-peer communications:} P2P communication in Metaverse guarantees direct machine-to-user or user-to-machine communication using DTs with no intermediary. 
\item \textit{Access privileges and trusted DTs data coordination:} Blockchain, as Metaverse infrastructure, provides accessibility to DTs data, which is easy to manage for the company coordinators. 
\item \textit{Enforcing transparency and accountability for DTs data:} Transparency is one of the most popular blockchain features and user-friendly properties in Metaverse. Additionally, accountability can alleviate regulatory issues for DT legitimacy and usage.  
\item \textit{Decentralized Infrastructure:} Metaverse provides decentralized infrastructure for DTs where all blockchain properties are supported. Therefore, applying DTs in Metaverse is a good and reliable choice. 
\end{itemize}

Fig. \ref{rel} outlines the suggested three-layer architecture of generating and applying DTs in Metaverse. In the following, three layers and workflows of the architecture are described. 

\subsection{Physical/Real World Layer}
Real-world users and components demand highly-accurate DTs of themselves in Metaverse. As mentioned, DTs are the best choice for this demand. 

As the name shows, the physical/real world layer contains users, things, and services (market places, healthcare centers, shops, entertainment, etc) in the physical world. For connecting to the digital world (called \textit{Metaverse} in this paper), users and company owners send their requests to the link layer and pay fees.

DTs can be designed as a wide world, and the biggest DT is the DT of the world. Popular or the most applicable DTs are industrial and business DTs. Additionally, people also like to have highly-accurate avatars in Metaverse. It, therefore, motivates people to use DTs in Metaverse.

\subsection{Link (User Interface) Layer}
It is suggested that the most critical layer of the architecture is the one that links the physical world to Metaverse. It consists of the two following sub-layers.

\subsubsection{Simulation/Migration}
It is the first sub-layer of the Link layer, where NFT generator services (e.g., Opensea, Rarible, etc) are present. In addition,  computers (e.g., 3D scanners, etc) and developers (e.g., programmers) work there to create digital versions of physical things (digital twins (DTs)).

Programmers and developers try to create highly-accurate, fully detailed DTs for presenting natural feelings in Metaverse for real-world users. Therefore, expert developers, large companies, and powerful computers are in this layer and compete to attract more clients.

After generating the DT, the service/user that created it submits the developed DT to the blockchain and pays for it. 

\subsubsection{Blockchain}
Submitted data, especially DTs, are ready to use after submission on the blockchain, and DApps and other services have access to them (it should be noted that the submission process is out of the scope of this paper - for more details on this topic, refer to \cite{bitcoin,ijns}).

Blockchain is assumed as the Metaverse infrastructure where miners, smart contracts, blockchain nodes, full nodes, and other components exist. Blockchain, as a distributed ledger, records all Metaverse correspondences and transactions. People (in the physical world) and users (in Metaverse) have access to the blockchain and can submit new transactions or content, and are also able to read and use the submitted content.

\subsection{Metaverse (Digital World) Layer}
The most attractive layer of the suggested architecture provides a 3D digital world (Metaverse). DTs are used in this layer by relying on blockchain and smart contracts. As with the physical world, all people, services, and things could be present in the Metaverse layer as DT or NFT, and they can enjoy the digital environment or resolve their problems. 

The DT-based digital world provides everything, including people's avatars, businesses, retail markets, and manufactories with all high-accuracy industrial things, satisfying daily requirements of regular people and managers cost-efficiently, remotely, and digitally. \\

\textit{Note: In \cite{mv2}, a three-layer architecture of Metaverse, with a different approach, was presented by Duan \textit{et al.}. It should be said that the proposed three-layer architecture in \cite{mv2} focuses on a general model of Metaverse. However, the three-layer architecture proposed in this study involves the Link layer with the aspect of applying DTs in Metaverse, not a general model.}

\section{Security and Privacy Challenges} \label{secchall}
This section first discusses the security aspects of using DTs in Metaverse. It then describes privacy issues (it should be highlighted/repeated that  the original definitions of the addressed security and privacy issues are not presented here; they are defined with the aspect of using DTs in Metaverse).

\subsection{Security}
Data security is generally defined as \underline{C}onfidentiality, \underline{I}ntegrity, and \underline{A}vailability of Data (CIA) \cite{sec1,sec2}. The three main aspects of security and some other security properties will now be discussed below.
\subsubsection{Confidentiality}
As in the physical world, people want to have secure and confidential transactions/correspondence. Confidentiality, as the security feature requested by users, should therefore be supportable as an optional feature based on users' demands. Similarly, confidentiality in Metaverse is better as an optional or mandatory feature for users who requested to use DTs.
\subsubsection{Integrity}
Submitted data, especially DTs, should not be changed along the submission process and after the submission. Used DTs should be the same DTs that have been created/submitted previously with no changes since DT details are critical to them, and many damages may occur after each change.
\subsubsection{Availability}
Users (in the real or digital world) demand that their DTs be available or accessible at all times and places. Moreover, company owners or service providers do not want to be out of service. These facts indicate that availability is a critical feature demanded by all types of users presented in the digital world where DTs play essential roles.
\subsubsection{Authentication}
From the past to the present, people/users have wanted to know their opposite parties completely or authenticate their validity. This is also true in Metaverse where the people/users in the digital world want to ensure the Metaverse-based services and the validity of used DTs. Therefore, mechanisms should be provided for proving the DT validity. 
\subsubsection{Central Management}
Decentralization is a popular and essential feature of the Metaverse. However, there is one person or a centralized group of authority on the background of most Metaverse-based services who manage users and components (e.g., DTs) in Metaverse. Therefore, providing Metaverse-based services with no central authority is a critical challenge. 
\subsubsection{Identity Management}
As with currently-in-use social networks, Metaverse can be a suitable infrastructure for criminals and users who misbehave. Therefore, identity management, including registration, revoking, and updating is a challenge in Metaverse (it should be noted that the phrase \textit{identity} is not assigned only to users. However, it is assumed as the digital identity of DTs, NFTs, or other digital entities). Submitting high-accuracy DTs, updating the submitted DTs, and revoking invalid DTs (DTs management) are important issues in Metaverse.
\subsubsection{Duplication Data}
Uniqueness is a valuable feature expected from the DTs. Copies or fake versions of DTs can be mistaken as the same valuable DTs. This issue (duplicating DTs to fake versions) is a challenge for DTs owners and can decrease value and validity of their DTs. Therefore, mechanisms should be provided to prevent DT duplication in Metaverses.
\subsubsection{Cyber Attack}
Although blockchain technology provides some security features (e.g., security against DoS and DDoS attacks, immutability, etc) for its applications, cyber-attacks are implementable on DTs in Metaverse. Protecting against common attacks is a primary necessity for DTs and the digital world.

\subsection{Privacy}
Privacy is a feature requested by people/users in the physical/digital world(s). However, privacy is not a bounded feature, and its bounds are determined based on users' demands and conditions \cite{pri1,pri2}. In the following, some privacy issues are described regarding using DTs in Metaverse.
\subsubsection{Users Privacy}
Users want to be safe and have privacy in both the physical and digital worlds. As mentioned above, privacy has no certain bounds. This paper, however, assumes anonymity and untraceability as two aspects of privacy. Anonymity refers to having secure pseudonyms, and untraceability is when no one can find links between users' activities. Users who use DTs in Metaverse demand this feature for safety.
\subsubsection{Trust}
Validity, authentication, and mutual authentication are necessary to create trust in users. However, providing them in a decentralized environment where no trusted third party or judgment mechanism is present is challenging. In other words, it is hard or almost  to provide trust by untrusted parties. Therefore, users who use DTs in Metaverse need to have a logical trade-off, based on themselves tastes, between faith in untrusted parties and their privacy.
\subsubsection{Assets Ownership Proof and Security}
Based on the value of DT and their applications, hackers and criminals want to steal their ownership. There must, therefore, be a reliable and secure mechanism to prove their ownership publicly provable, and no one can forge that.
\subsubsection{Ownership Transferring}
As with the discussed privacy factors, transferring the ownership of DTs needs the support of security and privacy aspects (e.g., confidentiality, anonymity, untraceability, etc) based on user demands.
\subsubsection{Money-laundering}
Transactions related to buying/selling DTs on Metaverse could be examples of money-laundering if privacy for users and confidentiality for financial transactions are provided.
\subsubsection{Conditional Privacy and Government Monitoring}
Conditional privacy is a prominent feature for governments and authorities who want to control communities. It means they, as authorities or judges, can break users' privacy. It is, however, computationally hard for invalid users (e.g., adversaries, malicious users, or hackers). In this case, privacy refers to DTs ownership, correspondences, and financial transactions related to them.

\section{Conclusion and Future Works} \label{conc}
This study first presented definitions of concepts, including blockchain, NFT, Metaverse, and DT. It then suggested and described a three-layer architecture that indicated the application of DTs in Metaverse. Finally, it discussed the security and privacy challenges of using DTs in Metaverse as the paper's most important section. 

In the following, some possible solutions for solving the discussed challenges and future work of research in this field will be described.
\begin{itemize}
\item \textit{Using blockchain-based security protocols supporting conditional privacy:} As with the real world and internet-based communication, users in the digital world or people in the Metaverse request privacy. However, privacy provides suitable environments for criminals, and users do not want that. Therefore, for controlling malicious users, conditional privacy \cite{f1} is a critical feature for Metavese-based services to prevent fraud and insecurity.
\item \textit{Using blockchain-based payment methods with no central authority:} The existence of a central authority is against the distributed implementation of Metaverse-based services. Therefore, for providing reliability in distributed payment systems, it is suggested to specialize payment methods with no central authority \cite{f2} for Metaverse.
\item \textit{Using blockchain-based data auditing protocols:} Regarding the value of DTs, protecting DTs' integrity in the digital world is a critical issue. So, using blockchain-based data auditing protocols \cite{f3}, specialized for Metaverse, for proving DTs integrity and preventing DTs duplication (uniqueness proof) is suggested.
\item \textit{Applying blockchain-based reporting protocols:} Similar to the real world, malicious users exist in the Metaverse. It, therefore, is suggested to use blockchain-based reporting protocols \cite{f4} in the Metaverse for detecting malicious users easier.
\item \textit{Applying zero-knowledge based cryptocurrencies:} As aforementioned, privacy is a user-demanded feature in the Metaverse. Therefore, for users who buy/sell DTs and NFTs, applying zero-knowledge-based cryptocurrencies \cite{f5} for providing anonymity and untraceability is suggested. 
\item \textit{Using blockchain-based anonymous authentication protocol:} All users want to authenticate the validity of their environments users along with anonymity \cite{f6}. Therefore, to provide mutual authentication in related-to-DTs' transactions and users' privacy, it is suggested to use blockchain-based anonymous authentication protocols in the Metaverse. 
\item \textit{Assigning rewards to users who help with the network maintenance:} The users who help with network maintenance should be motivated, and using methods assigning rewards to users \cite{f7} empower users' motivation for maintaining the network and increase network reliability.
\end{itemize}
It is clear that the discussed concepts, services, and technologies, including blockchain, NFT, Metaverse, and DT, are beyond the scope of this paper. Therefore, several fields remain for future studies and are not limited to the above items. \\

\textit{Conflict of Interest:} The authors received no financial support for the research and/or authorship of this article. The authors declare that they have no conflict of interest in the publication of this article.

\end{document}